\documentstyle[prb,aps,epsf,twocolumn]{revtex}
\begin{document}
\draft
\twocolumn[\hsize\textwidth\columnwidth\hsize\csname@twocolumnfalse%
\endcsname
\title{Static and dynamic image potential for tunneling into a
Luttinger liquid} 
\author{S.~H\"ugle, R.~Egger and H.~Grabert}
\address{Fakult\"at f\"ur Physik, Albert-Ludwigs-Universit\"at,\\
Hermann-Herder-Stra{\ss}e 3, D-79104 Freiburg, Germany}
\date{Date: \today}
\maketitle
\widetext
\begin{abstract}
We study electron tunneling from a tip or a lead 
into an interacting quantum wire described by Luttinger liquid theory.
Within a WKB-type approach, the Coulomb
interaction between the wire and the tunneling 
electrons, as well as the finite traversal time 
are taken into account.  Although the static image potential 
is only logarithmically suppressed against the bare
Coulomb interaction, the dynamic image potential
is not strong enough to alter power-law exponents
entering the tunneling density of states.
\end{abstract}
]
\narrowtext

One-dimensional (1D) quantum wires (QWs) are at the focal point of current
activities in condensed-matter physics.  Fabrication
advances in semiconductor heterostructures\cite{wire}
and carbon nanotubes\cite{tube} allow the systematic
study of phenomena arising only in one dimension. 
In particular, the physics of 1D nanowires is intimately connected
with the concept of a Luttinger liquid (LL),
exhibiting spin-charge separation, suppression of the 
tunneling density of states
(TDOS), and interaction dependent power-laws in transport
properties.\cite{voit,gogolin} 
For nanotubes, the theoretically predicted\cite{lltube1} 
LL behavior has been convincingly established
in several recent experiments.\cite{lltube2}
Of particular importance was the measurement 
of the TDOS power-law exponent $\alpha$, 
which provides information about the dimensionless
interaction strength parameter $g$ of the LL (where $0<g\leq 1$).
The TDOS can be obtained, e.g.~from the tunneling current
measured via weakly attached metallic leads.  Another possibility 
is to use a scanning tunneling microscopy (STM) tip. 
Apart from the resolution of atomic and electronic properties of individual
nanotubes successfully achieved in recent 
STM experiments,\cite{stmtube1,stmtube2,stmtube3} a detailed
analysis of nonlinear current-voltage curves in the spectroscopy
(STS) mode would allow to extract the TDOS exponent.

So far in all theoretical studies of tunneling into a 
LL, it has tacitly been assumed that the 
electrons tunneling into the correlated
fermion liquid do not modify the effective TDOS.
Conventional treatments\cite{eggert} employ
the tunneling Hamiltonian and thereby
assume the traversal time of tunneling\cite{landauer}
to be infinitely short. In addition, the interaction
between the tunneling electron and the LL electrons is 
neglected. Given these two assumptions, the TDOS is indeed an intrinsic
property of the LL with a power-law dependence on energy,
$\rho(\epsilon)\propto\epsilon^\alpha,$ for $\epsilon \to 0$,
so that $\alpha$ depends only on $g$.  (For clarity,
we focus on the case of bulk tunneling at zero temperature
throughout this paper.)  For an STS experiment,
such a calculation predicts that the measured differential
conductance is proportional to the TDOS,\cite{tersoff}
and hence the exponent $\alpha$ can be extracted from
experimental data.  A similar reasoning has been employed to understand
the data in Ref.~\onlinecite{lltube2}, where electron tunneling
between metallic leads and a nanotube was important.

Here we address the question of {\sl whether this measured
exponent really characterizes the unperturbed LL}, 
or whether it is affected by
the dynamics and charge of the tunneling particle.  To that end, we 
take into account correlations both within the QW and 
between the QW and the tunneling electron, as well
as a finite traversal time.  Our main findings are as follows:
Although the {\sl static} image potential experienced by
the tunneling electron is very strong, dynamical effects
turn out to be of crucial importance.  
Within the framework of a semiclassical theory 
related to Nazarov's tunnel junction theory,\cite{nazarov}  
the low-energy power-law exponent of the TDOS is
governed solely by intrinsic LL properties.  Thereby
we provide the {\sl a posteriori}\ justification for the (previously 
assumed) connection linking the (observable) value of $\alpha$ and the
interaction parameter $g$.  Since spin and charge are decoupled in a
LL, we study only the spinless single-channel case, with
the same conclusions applying to spin-$\frac12$ electrons  or
nanotubes. 

\begin{figure}
\epsfxsize=1.0\columnwidth
\epsffile{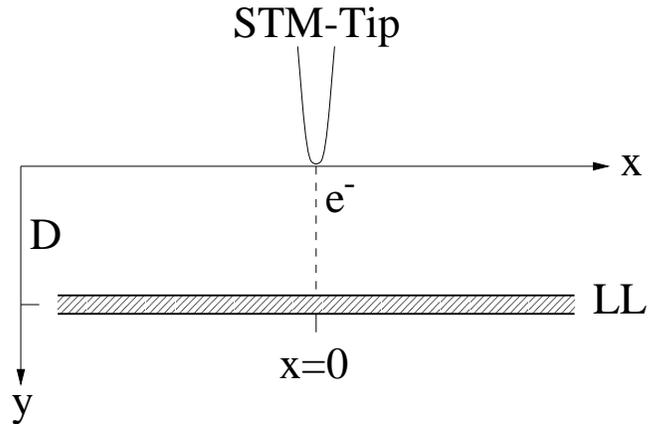}
\caption[]{\label{tunnelgeom} 
Tunneling from an STM tip into a Luttinger liquid (schematic).} 
\end{figure}

For the case of tunneling from an STM tip into
the QW, the relevant geometry is depicted in Figure \ref{tunnelgeom}.
We consider a clean and very long $(L\to \infty)$ QW,
and mainly focus on an interaction potential of the form 
\begin{equation}  \label{coul}
V(x,y)=\frac{1}{\sqrt{x^2+(D-y)^2 +a^2}}\;, 
\end{equation} 
where $a$ is the lattice constant of the QW. For $y=D$, this describes
the intra-wire interaction responsible for the LL state, while for $0<y<D$,
it gives the interaction between the wire electrons and the tunneling 
electron at $x=0$ (see below).  The one-sided Fourier transform of 
Eq.~(\ref{coul}) is
\begin{equation} \label{fourier}
\tilde V(q,y)=2K_0\left(|q|\sqrt{(D-y)^2+a^2}\right)\;,
\end{equation}
 with the modified Bessel function $K_0$. 
When using Eq.~(\ref{coul}), one neglects
or crudely approximates several important effects, 
e.g.~screening due to the tip, or the orbital structure of both tip and 
quantum wire (see, e.g.~Ref.~\onlinecite{stmtube3}).
However, our qualitative findings concerning the 
connection between $\alpha$ and $g$ are expected to hold
quite generally.  In addition to the model potential (\ref{coul}),
these findings can be made rigorous for a general class
of separable potentials.

Tunneling out of the tip (or lead) proceeds from the ground state
with energy $E_0'$ into the LL ground state with energy $E_0$,
and is usually described in terms of a simple tunneling Hamiltonian, 
\begin{equation}
H_T=T\Psi^\dagger(0)c_0+{\rm h.c.} \;,
\end{equation}
where $c_0$ annihilates an electron in a state at the center of the tip and
$\Psi^\dagger(0)$ creates an electron at $x=0$ in the LL. If the tunneling
matrix element $T$ is small, $H_T$ can be treated
as a perturbation to $H'+H$, where $H'$ describes the electrons in the
tip and $H$ denotes the Hamiltonian of the LL. More specifically, the
perturbative treatment is appropriate when the tunneling resistance 
is large compared to $h/e^2$. Then the 
tunneling rate $\Gamma$ can be calculated
with the aid of Fermi's golden rule.
Labeling states in the tip
by  $|\nu\rangle$ and states in the LL by $|n\rangle$, 
the tunneling rate is (we put $\hbar=1$)  
\[
\Gamma=2\pi T^2 \sum_{\nu,n}|\langle\nu,n|
\Psi^\dagger(0)c_0|0,0\rangle|^2 \delta(E'_0+E_0-E'_\nu-E_n) \;,
\]
where $|0,0\rangle$ denotes the ground state in the absence of tunneling.
Using the identity
\[
\delta(E_0'+E_0-E_\nu'-E_n)=\int\!d\epsilon\; \delta(E_0+\epsilon-E_n)
\delta(E_0'-\epsilon-E_\nu') \;,
\]
the rate may be written as 
\begin{equation} \label{trate}
\Gamma=2\pi T^2\int_0^{eV}\! 
d\epsilon \;\rho(\epsilon)\rho'(eV-\epsilon)  \;.
\end{equation}
Here the TDOS for adding an electron with energy $\mu+\epsilon$ to the LL is
\begin{equation} \label{rho}
\rho(\epsilon) = \sum_n |\langle
n|\Psi^\dagger(0)|0\rangle|^2\delta(E_0+\mu+\epsilon-E_n) \;,
\end{equation}
where the electrochemical potential
$\mu$ is the minimal energy required to add an electron. Furthermore,
\begin{equation}
\rho'(\epsilon') = \sum_\nu |\langle
\nu|c_0|0\rangle|^2\delta(E'_0-\mu'+\epsilon'-E'_\nu)
\end{equation}
is the DOS for removing an electron from the tip/lead
with energy $\mu'-\epsilon'$.
Now $\rho(\epsilon)$ and $\rho'(\epsilon)$ are nonvanishing only for
$\epsilon>0$, and $\mu'-\mu=eV$ determines the applied voltage.
With these definitions we readily obtain Eq.~(\ref{trate}).
In a metallic lead, the DOS $\rho'(\epsilon)$ is essentially
constant, while for an STM tip, we expect pronounced peaks reflecting
the discrete level structure.  In the latter case, Eq.~(\ref{trate}) 
reproduces the rate from an STM tip into a metal first obtained by
Tersoff and Hamann.\cite{tersoff}

To determine $\rho(\epsilon)$ explicitly,
we write 
\begin{eqnarray}
\rho(\epsilon) &=& \frac{{\rm Re}}{\pi}\int_0^\infty\!dt\; \sum_n \langle
0|\Psi(0)|n\rangle\langle n|\Psi^\dagger(0)|0\rangle
e^{i(E_0+\mu+\epsilon-E_n)t}\nonumber\\
\label{tdos} &=& \frac{{\rm Re}}{\pi}\int_0^\infty\!dt\; 
G_0(t) e^{i(\mu+\epsilon)t} \;,
\end{eqnarray}
with the single-electron Greens function ($t>0$)
for an electron at position $x=0$ in the ground state,
$G_0(t) = \langle\Psi(0,t)\Psi^\dagger(0,0)\rangle$.
To evaluate $G_0(t)$, standard bosonization methods\cite{voit,gogolin}
can be applied.  The kinetic part of the Hamiltonian is
\[
H_0=\frac{v_F}{2}\int dx 
\left[(\partial_x\vartheta)^2+(\partial_x\varphi)^2 \right]+\mu\hat{N}\;.
\]
Here $v_F$ is the Fermi velocity, $\hat N$ the
particle number operator, and $\vartheta(x)$ is conjugate to the phase field
$\varphi(x)$ describing the plasmon excitations in the wire. 
The intra-wire interaction part is
$H_V=\frac12\int\! dx\,dx'\; n(x) V(x-x',D) n(x')$ with the electron charge
density $n(x) = -e \pi^{-1/2} \partial_x\varphi$.
(The LL model appropriate for the low energy sector follows
by effectively using a local interaction, $V(x,D)=V_0\delta(x)$.)
By virtue of a Bogoliubov transformation, $H=H_0+H_V$ can easily
be diagonalized.  With bosonic
operators $b_q^{(\dagger)}$, the phase field $\varphi(x)$ reads\cite{gogolin}
\begin{equation} \label{7}
\varphi(x) = i\sum_{q\ne 0}\left(\frac{g(q)}{2L|q|}\right)^{1/2}
\exp(-iqx) \,
{\rm sgn}(q) \, [ b_q^\dagger+b_{-q} ]  \;, 
\end{equation}
with the $q$-dependent interaction parameter [where $g= g(q=2\pi/L)$]
\begin{equation} \label{9}
g(q)=[1+e^2\tilde V(q,D)/\pi v_F]^{-1/2}=g(-q)\;.
\end{equation}
We then arrive at
\begin{equation}
H=\sum_{q\ne 0}\omega_qb_q^\dagger b_q + \mu \hat{N} \;,
\end{equation}
with the plasmon dispersion relation $\omega_q=v_F|q|/g(q)$.
In terms of the chiral (right- or left-moving) phase fields ($p=R/L=\pm$),
\begin{equation}
\label{phi}
\phi_p(x)=[p\varphi(x)+\vartheta(x)]/\sqrt{4\pi} \;,
\end{equation} 
the bosonized electron operator at $x=0$ is
$\Psi(0,t)\propto \sum_{p=\pm} \exp[ 2\pi i\phi_p(0,t)]$, 
implying\cite{voit,gogolin}
\[
G_0(t)\propto t^{-(g+g^{-1})/2}\; e^{-i\mu t}  
\]
at long times. Hence one obtains the well-known exponent 
$\alpha=(g+g^{-1}-2)/2$  governing the bulk TDOS.

Let us now look at the {\sl static image potential}\ experienced 
by an electron with charge $-e$ held fixed at position $x=0$ and $0<y<D$
due to its interaction with the QW electrons,
$H_I = -e\int \!dx\;n(x)V(x,y)= -e\phi(y)$.
Using Eq.~(\ref{7}), we get the fluctuating field
\begin{equation} \label{12}
\phi(y,t) = \sum_{q\ne 0}\lambda_q(y) \left(b_q^\dagger(t)+b_{-q}(t)\right) 
\end{equation}
with couplings
$\lambda_q(y)=-e [g(q)|q|/2\pi L]^{1/2}\,\tilde V(q,y)$.
Next we shift the bosonic operators,\cite{persson} 
$B_q=b_q-e\lambda_q(y)/\omega_q$,
whence
\begin{equation}
H=\sum_{q\ne 0}\omega_qB_q^\dagger B_q+ \mu \hat{N}+V_{im}(y) \;.
\end{equation}
Here the static image potential
\begin{equation} \label{vim}
V_{im}(y)=-e^2\sum_{q\ne 0}\lambda^2_q(y)/\omega_q
\end{equation}
describes the energy gained by the plasmons relaxing to their
equilibrium state in the presence of the additional electron. 
For the unscreened interaction (\ref{coul}), one has
$g(q)=[1+\xi K_0(|q|a)]^{-1/2}$ with
the dimensionless parameter $\xi=2e^2/\pi v_F$. Then the 
image potential (\ref{vim}) for $(D-y)\gg a$ reads
\[
V_{im}(y)=-\frac{2e^2/\pi}{(D-y) \{\ln[(D-y)/a]+1/\xi\}} \;.
\]
Therefore the static image potential is {\sl only logarithmically suppressed}\
against the bare Coulomb interaction, and hence is very strong.

Next we turn to dynamical effects due to tunneling.
We envision the latter as penetration through a rectangular barrier
of width $D$. If the barrier is sufficiently thick and its
transparency low, the main contribution to the tunnel current comes from
electrons with momenta perpendicular to the QW.  Therefore
we effectively obtain a 1D Schr\"odinger equation
for the underbarrier motion $\psi(y,t)$ of the tunneling electron 
 ($0<y<D$),
\begin{equation} \label{16}
i \partial_t \psi(y,t)=[ -(2m)^{-1}\partial_y^2+\mu'+U-e\phi(y,t)]
\psi(y,t)\;,
\end{equation}
where $U$ is the work function of the tip/lead.
In the absence of $\phi(y,t)$, 
the solution for an electron at energy $\mu'$ reads 
$\psi(y,t) \propto e^{-i\mu' t-mvy}$, where
$v=\sqrt{2U/m}$ is an effective velocity related to the traversal time
$D/v$. Under the WKB approximation, the dominant effect of the
potential $\phi$ can be incorporated as additional phase factor,\cite{nazarov}
\begin{equation} \label{15}
\psi(y,t) \propto e^{-i\mu' t-mvy-i\theta(y,t)} \;.
\end{equation} 
Linearizing the resulting WKB equation 
gives for $|\phi(y,t)|\ll U$:
\begin{equation} \label{17}
\partial_t\theta+iv\partial_y\theta=-e\phi(y,t) \;,
\end{equation}
supplemented by the boundary condition $\theta(0,t)=0$. This
equation can be solved separately for each bosonic mode 
using the ansatz 
\[
\theta(y,t)=\sum_{q\ne 0}\left[ w(y,\omega_q) b_{-q}(t)+ \tilde{w} (y,\omega_q)
b_q^\dagger(t) \right]\;. 
\]
$ $From Eq.~(\ref{17}) and $b_q(t)=b_qe^{-i\omega_qt}$, we obtain
$\tilde{w}(y,\omega_q)=w(y,-\omega_q)$ and 
\[
(v \partial_y - \omega_q ) w(y,\omega_q) = ie\lambda_q(y) \;,
\]
which can easily be solved. For $y=D$, we finally get
\begin{equation}
\theta(t)=\sum_{q\ne 0}\left[ w(-\omega_q) b_q^\dagger(t) +
w(\omega_q)b_{-q}(t) \right] \;, 
\end{equation}
with 
\begin{equation}
w(\omega_q)=(ie/v) \int_0^D dy\, \lambda_q(y)\exp[\omega_q(D-y)/v] \;.
\end{equation}
Because of the associated dynamic image potential, the electron wave
function acquires the phase factor $\exp[-i\theta(t)]$ 
during the tunneling process. 

\begin{figure}
\epsfxsize=1.0\columnwidth
\epsffile{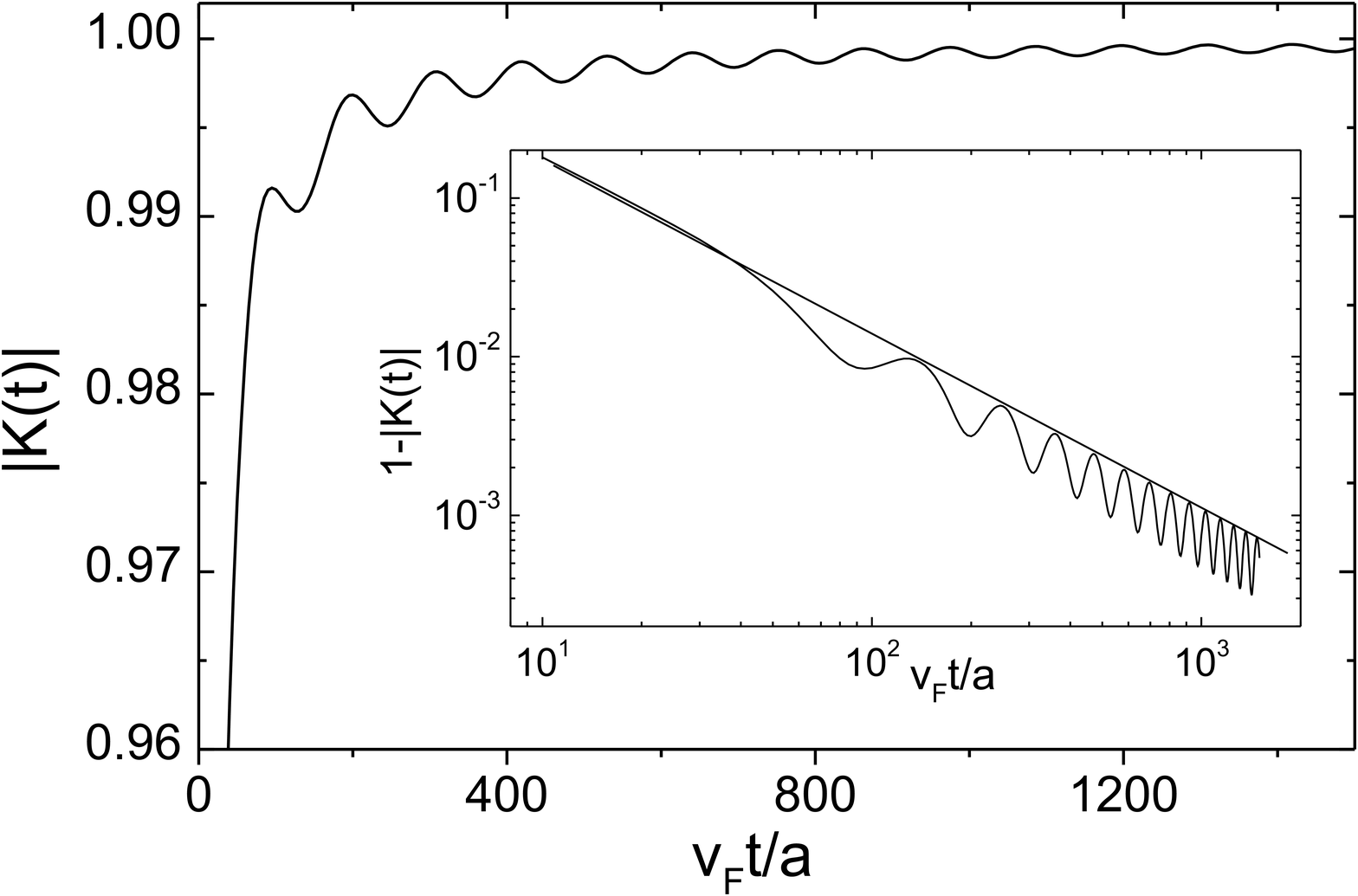}
\caption[]{\label{pb2} 
Time-dependent correction factor $|K(t)|$ for the unscreened interaction
(\ref{coul}). The parameter values are chosen as $\xi=1.72$, $D/a=3$, and
$v/v_F=1.48$. These values should be appropriate for nanotubes, where
$U\approx4$~eV and $v_F\approx8\times 10^5$ m/sec. The inset shows the
decrease of the oscillation amplitude using double-logarithmic scales. The
straight line is a guide to the eye only.}  
\end{figure}

This effect can be properly incorporated by a
modification of the tunneling Hamiltonian,
\begin{equation}
\tilde H_T=T\Psi^\dagger(0)e^{-i\theta}c_0+{\rm h.c.}\; .
\end{equation}
The Greens function determining  the effective 
TDOS is now given by 
\begin{equation}
 G(t)=\langle \exp[i\theta^\dagger(t)]\Psi(0,t)\Psi^\dagger(0,0)
\exp[-i\theta(0)]\rangle\;.
\end{equation}
Putting $G(t)=G_0(t)K(t)$,
the contribution of the dynamic image potential then gives rise to 
the factor 
\begin{equation} \label{28}
K(t)=\exp[C_1(t)+C_2(t)] \;,
\end{equation}
where we introduce the functions
\begin{eqnarray*}
C_1(t) &=& -\frac{1}{2}\langle\theta^{\dagger 2}(t)+\theta^2(0)
-2\theta^\dagger(t) \theta(0)\rangle \;,\\
C_2(t) &=& 2\pi\left[\langle\phi_p(t)\theta(0)\rangle+
\langle\theta^\dagger(t)\phi_p(0)\rangle \right. \\
& &\left.-\langle\theta^\dagger(t)\phi_p(t)\rangle- 
\langle\phi_p(0)\theta(0)\rangle\right],
\end{eqnarray*}
where $C_2$ is independent of $p=\pm$.
Doing the Gaussian averages yields
\begin{eqnarray} \label{31}
C_1(t) &=& -\sum_{q\ne 0} [w(\omega_q)w(-\omega_q)+w^2(-\omega_q) 
\exp(-i\omega_q t) ] \;,
\\ \label{32}
C_2(t) &=& i (2\pi/L)^{1/2}
\sum_{q\ne 0}\frac{w(-\omega_q)}{\sqrt{g(q)|q|}} [ 1-\exp(-i\omega_qt) ].
\end{eqnarray}
Since we are interested in the power-law exponent governing
the TDOS, we focus on the time-dependent parts of 
Eqs.~(\ref{31}) and (\ref{32}), and do not explicitly compute the 
prefactor.
For the interaction (\ref{coul}), numerical calculation of $K(t)$ 
 gives the result shown in Fig.~\ref{pb2} which is well
approximated by $|K(t)| =  1+A(t)\cos(\Omega t)$ for long 
times, with oscillation frequency $\Omega$. The amplitude 
decays according to $A(t)\propto t^{-\beta}$ with 
$\beta \approx 1.15$. This result is insensitive to the precise parameter
values taken for $\xi, D/a$, and $v/v_F$. 
As a consequence, the {\sl power-law exponent}\ $\alpha$
of the TDOS for small energy $\epsilon$ remains {\sl unchanged
by the dynamic image potential}.
Hence one can indeed obtain the LL parameter $g$ from
a measurement of $\alpha$.

This finding can be inferred analytically for a
class of separable interaction potentials of the form
\begin{equation}
V(x,y)=V_0\delta(x)f(y) \;,
\end{equation}
where $f(y)$ is an arbitrary function with
$f(D)=1$. In this case, we get $\tilde V(q,y)=V_0f(y)$, 
leading to $g(q)=g$. 
The time-dependent parts of Eqs.~(\ref{31}) and (\ref{32})
read
\begin{eqnarray*}
C_1(t) &\propto& \int_0^\infty\!dq\;e^{-iv_F qt/g}q\left[\int_0^D\!dy\;f(y)
e^{-v_F(D-y)q/gv}\right]^2 \;,\\
C_2(t) &\propto& \int_0^\infty\!dq\;e^{-iv_F qt/g}\left[\int_0^D\!dy\;f(y)
e^{-v_F(D-y)q/gv}\right]\;.
\end{eqnarray*}
The asymptotic long-time behavior of $C_{1,2}(t)$
can be accurately calculated in stationary-phase approximation.
We find that $C_1(t)$ decays faster than $1/t$, while
$C_2(t) \propto 1/t$.
Therefore, from Eq.~(\ref{28}),
the TDOS exponent for small $\epsilon$ remains unchanged.
We expect this result to be correct and generic for arbitrary physically
relevant interaction potentials.

We conclude by summarizing our results. 
We have presented a simple theory of electron
tunneling from a tip or a lead into a strongly 
correlated 1D metal, explicitly incorporating 
the finite traversal time and the dynamic response
of the correlated metal to the incoming electron. We have solved
this problem within a WKB-type approximation for different 
interaction potentials.
Despite the presence of a strong static image potential,
the power-law exponent entering the tunneling density 
of states is not affected by these effects, but  completely
determined by the correlation strength in the 1D metal.

We acknowledge financial support by the Deutsche Forschungsgemeinschaft
under the Gerhard-Hess program.

\begin{references}

\bibitem{wire}
{S.~Tarucha, T.~Honda, and T.~Saku}, Solid State Comm. {\bf {94}},  413
  (1995); {O.~M.~Auslaender, A.~Yacoby, R.~de~Picciotto, K.~W.~Baldwin,
L.~N.~Pfeiffer, and K.~W.~West}, 
Phys. Rev. Lett. {\bf 84}, 1764 (2000). 

\bibitem{tube}
C.~Dekker, Physics Today {\bf 52}, 22 (1999).

\bibitem{voit}
{J.~Voit}, Rep.~Progr.~Phys. {\bf {57}},  977  (1995).

\bibitem{gogolin}
{A.~O.~Gogolin, A.~A.~Nersesyan, and A.~M.~Tsvelik}, {\sl Bosonization and
Strongly Correlated Systems} (Cambridge University Press, 1998).

\bibitem{lltube1}
{R.~Egger and A.~O.~Gogolin}, Phys.~Rev.~Lett. {\bf {79}},  5082  (1997);
{C.~L.~Kane, L.~Balents and M.~P.~A.~Fisher}, {\sl ibid.} {\bf {79}},
5086  (1997).

\bibitem{lltube2}
{M.~Bockrath, D.~H.~Cobden, J.~Lu, A.~G.~Rinzler, R.~E.~Smalley, L.~Balents,
and P.~L.~McEuen}, Nature {\bf {397}},  598  (1999);
{Z.~Yao, H.~W.~J.~Postma, L.~Balents, and C.~Dekker}, 
{\sl ibid.} {\bf 402}, 273 (1999).

\bibitem{stmtube1}
{J.~W.~G.~Wild\"oer, L.~C.~Venema, 
A.~G.~Rinzler, R.~E.~Smalley and C.~Dekker},
 Nature {\bf {391}},  59  (1998);
{T.~W.~Odom, J.~L.~Huang, P.~Kim and C.~M.~Lieber},
{\sl ibid.} {\bf {391}},  62 (1998).

\bibitem{stmtube2}
{L.~C.~Venema, J.~W.~G.~Wild\"oer, 
S.~J.~Tans, J.~W.~Janssen, H.~Tuinstra, 
L.~P.~Kouwenhoven and C.~Dekker}, Science {\bf {283}},  52  (1999).

\bibitem{stmtube3}
{C.~L.~Kane and E.~J.~Mele}, Phys.~Rev.~B {\bf {59}},  R12759  (1999).
See also {W.~Clauss, D.~J.~Bergeron, M.~Freitag, C.~L.~Kane,
E.~J.~Mele, and A.~T.~Johnson}, Europhys. Lett. {\bf 47}, 601 (1999).

\bibitem{eggert}
See, e.g., {S.~Eggert}, preprint cond-mat/9909001.

\bibitem{landauer} R.~Landauer and Th.~Martin,
Rev.~Mod.~Phys. {\bf 66}, 217 (1994).

\bibitem{tersoff}
{J.~Tersoff and D.~R.~Hamann}, Phys.~Rev.~Lett. {\bf {50}},  1998  (1983);
Phys.~Rev.~B {\bf {31}},  805  (1985).

\bibitem{nazarov}
{Yu.~V.~Nazarov}, Phys.~Rev.~B {\bf {43}},  6220  (1991).

\bibitem{persson}
{B.~N.~J.~Persson and A.~Baratoff}, Phys.~Rev.~B {\bf {38}},  9616  (1988).

\end{references}
\end{document}